\documentclass[12pt,a4paper]{article}

\usepackage{amsmath,amssymb,amsthm,amsfonts}
\usepackage{graphicx}
\usepackage[margin=1.0in,top=1.0in]{geometry}
\usepackage{setspace}
\usepackage{booktabs}
\usepackage{array}
\usepackage{tabularx}
\usepackage{longtable}
\usepackage{multirow}
\usepackage{makecell}
\usepackage{enumitem}
\usepackage{xcolor}
\usepackage{tcolorbox}
\usepackage{tikz}
\usetikzlibrary{positioning,calc,shapes.geometric,arrows.meta,decorations.pathreplacing}
\usepackage{hyperref}
\usepackage[utf8]{inputenc}
\usepackage[english]{babel}
\usepackage[autostyle,english=american]{csquotes}
\usepackage{caption}
\usepackage{subcaption}
\usepackage{titlesec}
\usepackage{fancyhdr}
\usepackage{microtype}
\usepackage{natbib}
\usepackage{float}
\usepackage{parskip}
\usepackage{threeparttable}
\usepackage{placeins}

\hypersetup{
  colorlinks=true,
  linkcolor=blue!60!black,
  citecolor=blue!60!black,
  urlcolor=blue!60!black
}

\definecolor{modA}{RGB}{70,130,180}
\definecolor{modB}{RGB}{60,160,80}
\definecolor{modC}{RGB}{210,120,30}
\definecolor{modD}{RGB}{160,50,130}
\definecolor{framegray}{RGB}{245,245,248}
\definecolor{ruleblue}{RGB}{0,51,102}

\tcbuselibrary{breakable,skins}
\newtcolorbox{designbox}[1]{
  colback=framegray,colframe=ruleblue,
  title={\small\textbf{#1}},breakable,
  fonttitle=\bfseries\small
}

\title{\large A Discipline-Agnostic AI Literacy Course for Academic Research:\\
Architecture, Pedagogy, and Implementation\\[0.6em]}

\author{
  Gideon K.\ Gogovi\\[0.3em]
  \small Department of Biostatistics and Health Data Science\\
  \small Lehigh University, Bethlehem, PA, USA\\[0.2em]
  \small \texttt{gig323@lehigh.edu}
}

\date{}

\begin{document}

\maketitle

\begin{abstract}
\noindent
The rapid integration of generative AI into academic workflows demands curricula that equip students not only with tool proficiency but with the critical judgment to use those tools responsibly in scholarly work. Existing offerings cluster around two inadequate poles: technical AI development courses serving narrow specialist audiences, and brief general-literacy interventions that cannot develop the sustained, practice-based competencies rigorous research requires. This paper reports the design, theoretical rationale, and implementation of BSTA~495/395: \textit{Getting Started with AI-Assisted Research}, developed and delivered at Lehigh University (Spring 2026). The course addresses an underserved gap: the competencies required for rigorous AI-assisted literature review. Its architecture organizes instruction into four sequential modules aligned with the cognitive demands of that task: comprehension of individual papers, construction and validation of knowledge taxonomies, identification of research gaps, and synthesis and production of complete literature reviews. Each module embeds an explicit verification discipline and standardized AI attribution practice. Prerequisite-free and discipline-agnostic, the course enrolls upper-level undergraduates and graduate students across all fields with differentiated assessment expectations. Pre- and post-course survey data from the inaugural offering indicate substantial self-reported confidence
gains, with the largest in hallucination detection ($d = +1.45$), responsible AI use ($d = +1.33$), and AI attribution practice
($d = +2.40$), consistent with the course's design emphasis. The course constitutes a replicable model for the emerging genre of AI research literacy curricula.\\[0.4em]

\noindent\textbf{Keywords:} AI literacy; curriculum design; literature
review pedagogy; generative AI; scaffolded learning; AI ethics;
prompt engineering
\end{abstract}

\section{Introduction}
\label{sec:intro}

Higher education institutions face a curricular challenge with few
precedents. Generative artificial intelligence tools have diffused into academic workflows faster than curricula can adapt, producing a student population that uses AI extensively but without the conceptual frameworks or critical habits of mind to use it well
\cite{cotton2024chatting,lund2023chatting}. The consequences are already visible: shallow engagement with primary literature as students offshore cognitive tasks that build foundational skills, and integrity violations as AI-generated content enters assessments without adequate verification or attribution \cite{milano2023large, stokel2023chatgpt}.\\
The problem, importantly, is not AI use per se. A growing body of evidence supports the view that well-deployed AI tools can function as effective cognitive scaffolds, accelerating comprehension, reducing access barriers to technical literature, and enabling researchers to process larger bodies of evidence than unassisted reading permits \cite{thirunavukarasu2023large,topol2019high}. The problem is
\textit{unsupported} AI use: deployment without understanding of tool limitations, without systematic verification practice, and without the scholarly judgment to distinguish what AI tools do reliably from what they do poorly or wrong.\\
Academic literature review is a domain where this gap is particularly consequential. Competent literature review requires skills spanning the full arc of scholarly work: evaluating individual studies, organizing knowledge across large document collections, identifying what is unknown, formulating researchable questions, and synthesizing evidence into coherent arguments. AI tools can assist at every stage of this arc, but each stage also presents characteristic failure modes (from hallucinated citations in comprehension phases to spurious pattern claims in synthesis) that require stage-specific critical skills to detect and correct
\cite{ji2023survey,alkaissi2023artificial}. A student who can use AI tools without knowing how they fail is not AI-literate; they are AI-dependent.\\
To address this gap, BSTA~495/395: \textit{Getting Started with AI-Assisted Research} was developed as a 13-week course at Lehigh University designed to cultivate AI research literacy as a coherent, teachable competency. The course's central argument is that responsible AI-assisted research requires four integrated skill clusters: \textit{technical proficiency} (knowing how to operate AI tools and craft effective prompts), \textit{critical evaluation} (knowing when AI outputs are reliable and when they require verification or correction), \textit{ethical practice} (knowing how to attribute AI contributions and maintain authorial integrity), and \textit{methodological judgment} (knowing which research tasks are appropriate for AI delegation and which require non-delegable human expertise). These four clusters are not independent competencies to be developed in sequence; they are mutually reinforcing and are developed in parallel throughout the course.\\
This paper has three goals. First, it documents the course's design in sufficient detail to enable replication at other institutions. Second, it articulates the theoretical and empirical rationale for key instructional design decisions, connecting them to relevant literature in pedagogy, AI literacy, and research methods education. Third, it proposes the course as a design model for the emerging genre of discipline-agnostic AI research literacy curricula.\\
This paper is a curriculum design paper in the tradition of design-based research in education \cite{brown1992design,collins1992toward}: it reports on a designed artifact, articulates the principles that motivated its design, and presents both implementation observations and preliminary outcome evidence that bear on the validity of those principles. Pre- and post-course survey data from the Spring 2026 cohort are reported in Section~\ref{sec:outcomes}; formal inferential analysis and a dedicated outcomes study are ongoing.\\
The paper proceeds as follows. Section~\ref{sec:background} situates the course within the existing landscape of AI education in higher education and presents the theoretical framework. Section~\ref{sec:design} describes the course architecture and the design decisions that produced it. Section~\ref{sec:modules} presents the four modules as implementations of coherent instructional principles. Section~\ref{sec:assessment} describes the assessment architecture and ethics infrastructure. Section~\ref{sec:lessons} presents implementation observations from the inaugural offering. Section~\ref{sec:outcomes} reports preliminary outcome evidence from pre- and post-course surveys. Section~\ref{sec:discussion} discusses implications and limitations. Section~\ref{sec:conclusion} offers recommendations for institutions developing comparable curricula.

\section{Background and Theoretical Framework}
\label{sec:background}

\subsection{Existing Approaches and Their Limitations}

AI-related instruction in higher education currently takes three broad forms, none of which adequately addresses the need for AI research literacy.\\
\textit{Technical AI education} (courses in machine learning, deep
learning, natural language processing, and related fields) develops
sophisticated understanding of how AI systems work but serves a narrow audience of technically oriented students and does not address the distinct problem of how researchers across disciplines should use AI tools as consumers rather than developers \cite{long2020ai}. The competencies this form of education develops are necessary preconditions for building AI systems, not sufficient conditions for using them rigorously in research.\\
\textit{General AI literacy} instruction typically takes the form of short modules or workshops embedded in introductory courses that introduce AI concepts, raise awareness of ethical risks, and provide basic orientation to tool use. While valuable as first exposure, such brief interventions cannot develop the scaffolded, practice-based competencies that sustained, rigorous AI-assisted research requires \cite{ng2021conceptualizing}. A student who has attended a workshop on AI risks has not thereby acquired the verification discipline to use AI tools safely in a literature review.\\
\textit{AI tool tutorials} focus on the mechanics of specific platforms without developing the underlying critical skills that make tools useful or safe across a changing tool landscape. This approach produces platform-specific proficiency that degrades rapidly as tools evolve and fails to transfer to novel contexts \cite{luckin2022empowering}. Knowing how to operate ChatGPT is not equivalent to knowing how to evaluate its
output.\\
What is absent from existing offerings is a course that integrates
technical AI proficiency, critical evaluation competencies, ethical
practice, and research methodology into a coherent semester-length
curriculum accessible to students across all fields. BSTA~495/395 was designed to fill this specific gap.\\
A recent and important addition to this landscape is \cite{shehu2026understanding}, who describes UNIV~182 at George Mason University, a prerequisite-free semester-long course in which non-major undergraduates reach the higher-order levels of Bloom's revised taxonomy through hands-on classifier construction, LLM probing, and team-based artifact design. Shehu's course demonstrates that technical depth and broad accessibility can coexist when scaffolding is designed to support both, and provides artifact-based evidence of reasoning progression from descriptive to design-level competency. BSTA~495/395 differs from this model in its organizing focus: where UNIV~182 centers the technical
pipeline of AI system construction, BSTA~495/395 centers the epistemological demands of academic literature review as the primary site of AI-assisted scholarly work. The competencies these courses develop are complementary rather than competing, and together they illustrate that prerequisite-free AI literacy instruction can be organized around meaningfully different research and professional workflows.

\subsection{The Specific Challenge of AI-Assisted Literature Review}

Literature review is a particularly important target for AI literacy
development for two reasons. First, it is a competency expected of
virtually all graduate students and increasingly of advanced undergraduates across disciplines, yet it is rarely explicitly taught as a standalone course \cite{machi2009literature,baltes2024teaching}. Students are typically expected to develop literature review skills through apprenticeship and feedback on research papers, a process poorly calibrated to an environment in which AI tools are now ubiquitous.\\
Second, the introduction of AI tools into literature review practice
creates risks that are qualitatively distinct from those in other research tasks. AI hallucinations in research contexts are often field-appropriate and internally consistent, making them difficult to detect without deep disciplinary knowledge. AI synthesis tends toward false consensus, smoothing genuine scholarly disagreements. And AI-generated prose can substitute for the deep engagement with evidence that produces genuine scholarly understanding, producing apparent competence that conceals its own shallowness. A course designed specifically for AI-assisted literature review must address these risks systematically, not as cautionary warnings but as subjects of analytical study.

\subsection{Scaffolded Learning and the Zone of Proximal Development}

The course's foundational pedagogical framework draws on Vygotsky's
concept of the zone of proximal development \cite{vygotsky1978mind}: the space between what a learner can accomplish independently and what they can accomplish with expert support. Well-designed scaffolding enables learners to operate in this zone, engaging with tasks beyond their current independent capability while building the competencies that will eventually make external support unnecessary.\\
Applied to AI-assisted research, this framework yields a clear design principle: AI tools should scaffold tasks of appropriate complexity while the student simultaneously develops the independent analytical skills that will ultimately allow competent work without AI support. The central pedagogical risk is what Wood, Bruner, and Ross \cite{wood1976role} termed scaffold dependency: using AI in ways that substitute for skill development rather than accelerating it. A student who asks AI to summarize a paper rather than reading it acquires neither comprehension of that paper nor the skill of efficient scholarly reading.\\
The course addresses this risk through a design principle I term
\textit{verified engagement}: students must be able to independently
explain, justify, and defend every claim they submit, regardless of how AI contributed to its production. This requirement operationalizes the scaffolding principle by ensuring that AI assistance produces understanding, not merely output.\\
This principle is independently instantiated in
\cite{shehu2026understanding}'s cumulative portfolio design, where each assessment builds the competencies required by the next, producing documented progression toward higher-order reasoning without prior technical preparation.

\subsection{Responsible Reliance Theory}

The course draws on the concept of responsible reliance as developed in automation and human factors research
\cite{parasuraman2008situation,lee2004trust} and extends it to academic research practice. Responsible reliance involves calibrating trust in automated systems to their actual reliability in specific task contexts: neither over-relying (accepting outputs uncritically) nor under-relying (rejecting useful assistance reflexively). Appropriate reliance in academic research requires knowing which AI capabilities are reliable and which are not, and adjusting verification effort accordingly.\\
The practical implication is that verification requirements should be calibrated to the empirical failure profile of each tool and task type. Citation generation by general-purpose large language models (LLMs) requires individual verification of every reference, because hallucination rates in this task are high. Structured extraction by index-backed platforms with direct paper access requires only targeted spot-checking, because hallucination rates are substantially lower when models retrieve from documents rather than generate from parametric memory. Students who apply uniform verification effort regardless of tool and task are neither efficiently using their time nor developing the judgment that expert AI use requires.\\
To illustrate: citation generation by a general-purpose LLM such
as ChatGPT or Claude draws entirely on parametric memory and carries a well-documented hallucination rate in the range of 30--50\% for specific bibliographic details \cite{alkaissi2023artificial,ji2023survey}. Full individual verification of every generated reference is therefore the appropriate default. By contrast, structured extraction by an index-backed platform such as Elicit or Semantic Scholar, where the model retrieves directly from an ingested paper rather than generating from memory, produces far fewer factual errors; targeted spot-checking of a 25\% stratified sample is sufficient verification for most research purposes. Students who understand this calibration principle can allocate their verification effort efficiently rather than uniformly, developing the task-specific judgment that distinguishes expert from novice AI use \cite{lee2004trust}. Recent human-AI collaboration research has extended the responsible reliance framework to generative AI settings, confirming that appropriate trust calibration, neither over-reliance nor reflexive rejection, is a learnable and teachable
competency \cite{li2024developing}.

\subsection{Threshold Concepts in Research Methods}

Meyer and Land's threshold concepts framework \cite{meyer2003threshold} identifies certain ideas as transformative entry points to disciplinary understanding: concepts whose mastery fundamentally reorients how learners see and engage with their field. In research methods education, classic threshold concepts include the epistemological distinction between correlation and causation, and the idea that evidence quality exists on a continuum that requires judgment rather than rule-following to navigate.\\
I argue that \textit{AI as cognitive scaffold rather than cognitive replacement} functions as a threshold concept in AI research literacy. Students who have genuinely crossed this threshold understand that AI assistance is legitimate to the extent that it amplifies their own analytical capacity, not to the extent that it substitutes for it. This understanding transforms every AI interaction: the evaluative question shifts from ``what did the AI produce?'' to ``what have I learned and can
I verify?'' 
Applying this framework requires demonstrating that the candidate
concept satisfies Meyer and Land's five diagnostic criteria
\cite{meyer2003threshold}: it must be \emph{transformative}
(reorienting how the learner engages with the entire domain),
\emph{irreversible} (difficult to unknow once genuinely understood),
\emph{integrative} (revealing previously hidden connections across
the field), \emph{bounded} (specific enough to be learnable), and
\emph{troublesome} (counterintuitive or resistant to surface
engagement). The concept of AI as cognitive scaffold rather than cognitive replacement satisfies each criterion. It is \emph{transformative}: students who have crossed it evaluate every AI interaction differently, shifting from ``what did the AI produce?" to ``what have I learned and can I independently verify?" It is \emph{irreversible}: researchers who have genuinely
internalized the distinction between AI-assisted and AI-replaced cognition cannot comfortably revert to uncritical AI delegation. It is \emph{integrative}: it connects prompt engineering, verification practice, attribution, and authorial ownership into a coherent stance rather than a list of separate rules. It is \emph{bounded}: it applies specifically to the use of AI tools in scholarly work rather than to AI literacy in its broadest sense. And it is \emph{troublesome}: students who have been conditioned to treat efficient outputs as the goal of any tool find the demand for independent verifiable understanding counterintuitive, even
burdensome, before crossing the threshold. The course is organized around progression toward and through this threshold concept, using each module's increasingly demanding AI tasks to force confrontation with the distinction between AI-assisted and AI-replaced cognition. Recent systematic reviews of the threshold concepts literature confirm the framework's continued relevance to technology-mediated learning contexts and its utility for designing instruction around transformative rather than merely additive competency development \cite{brown2022we}.

\section{Course Architecture and Design Decisions}
\label{sec:design}

\subsection{Course Identity and Structural Positioning}

BSTA~495/395: \textit{Getting Started with AI-Assisted Research} was
designed as a 13-week semester-length elective at Lehigh University,
offered simultaneously at undergraduate (395) and graduate (495) levels with shared instruction and differentiated assessment expectations. The course carries no prerequisites and enrolls students across all disciplines, a structural decision grounded in the premise that AI research literacy is a general scholarly competency rather than a field-specific one. A biostatistics course prefix was chosen for administrative reasons; the content is emphatically discipline-agnostic.\\
Weekly class meetings run for 2 hours and 40 minutes in a single extended session. This scheduling choice was deliberate: meaningful hands-on AI work requires sustained engagement that 50-minute periods cannot accommodate. Each session is organized in three phases: didactic instruction (approximately 45 minutes), guided hands-on laboratory activity (approximately 75 minutes), and structured debrief and synthesis (approximately 20 minutes), producing a rhythm that consistently moves between conceptual framing and empirical practice.

\subsection{The Four-Module Architecture}
\label{subsec:fourmod}

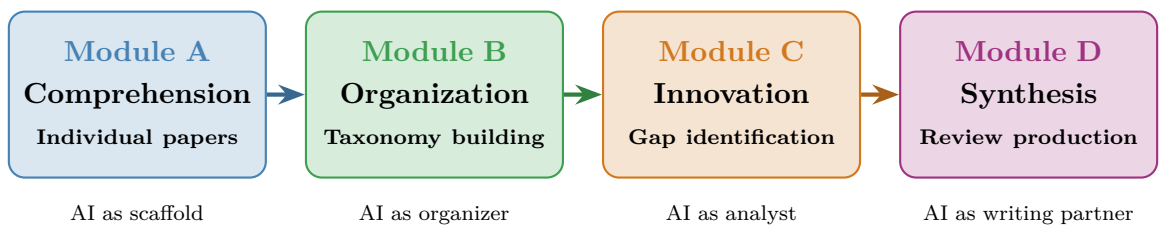
\begin{figure}[htbp]
\centering
\begin{tikzpicture}[
  node distance=0.5cm,
  mod/.style={rectangle, rounded corners=6pt, minimum width=3.3cm,
              minimum height=2.2cm, text width=3.1cm, align=center,
              font=\small\bfseries, draw, thick},
  arr/.style={-{Stealth[length=3.5mm]}, ultra thick}
]
  \node[mod, fill=modA!20, draw=modA] (A)
    {\textcolor{modA}{Module A}\\[2pt]{\small Comprehension}\\[2pt]
     {\scriptsize Individual papers}};
  \node[mod, fill=modB!20, draw=modB, right=of A] (B)
    {\textcolor{modB}{Module B}\\[2pt]{\small Organization}\\[2pt]
     {\scriptsize Taxonomy building}};
  \node[mod, fill=modC!20, draw=modC, right=of B] (C)
    {\textcolor{modC}{Module C}\\[2pt]{\small Innovation}\\[2pt]
     {\scriptsize Gap identification}};
  \node[mod, fill=modD!20, draw=modD, right=of C] (D)
    {\textcolor{modD}{Module D}\\[2pt]{\small Synthesis}\\[2pt]
     {\scriptsize Review production}};
  \draw[arr, color=modA!80!black] (A) -- (B);
  \draw[arr, color=modB!80!black] (B) -- (C);
  \draw[arr, color=modC!80!black] (C) -- (D);
  \node[font=\scriptsize, text width=3.2cm, align=center, below=0.2cm of A]
    {AI as scaffold};
  \node[font=\scriptsize, text width=3.2cm, align=center, below=0.2cm of B]
    {AI as organizer};
  \node[font=\scriptsize, text width=3.2cm, align=center, below=0.2cm of C]
    {AI as analyst};
  \node[font=\scriptsize, text width=3.2cm, align=center, below=0.2cm of D]
    {AI as writing partner};
\end{tikzpicture}
\caption{Four-module course architecture. Modules build sequentially, with each module's output serving as input to the next.}
\label{fig:modules}
\end{figure}

The course is organized into four sequential modules
(Figure~\ref{fig:modules}) that map directly onto the cognitive demands of literature review, scaling in unit of analysis from individual papers to complete synthetic arguments.

\FloatBarrier

This architecture embodies a core instructional design principle: the cognitive demands that define each module are qualitatively different, not merely more complex versions of the same demand. Comprehension (Module~A) requires extraction and evaluation of individual arguments. Organization (Module~B) requires pattern recognition and classification across a collection. Innovation (Module~C) requires evaluative judgment about the boundaries and lacunae of a field. Synthesis (Module~D) requires argumentation that integrates all prior work into an original scholarly position. AI tools play structurally different roles at each stage, and students encounter, analyze, and manage those different roles explicitly.\\
A second principle embedded in this architecture is the \textit{cumulative input design}: the primary output of each module (a structured reading log, a validated knowledge taxonomy, or a gap analysis report) serves as the primary input to the next module. This ensures that each module's skills are exercised not only in their immediate context but also in service of a downstream product, reinforcing the understanding that literature review is an integrated process rather than a sequence of separable tasks.

\subsection{Tool-Agnostic Design}

The AI landscape changes too rapidly for platform-specific training to remain valuable across a student's academic career. The course therefore develops transferable competencies (effective prompt construction, output verification, systematic bias detection, and attribution documentation) that apply regardless of which specific tools students use in any given period. In practice, instruction uses five platforms as vehicles for developing these competencies: ChatGPT (OpenAI) and Claude (Anthropic) as general-purpose LLM platforms, and Elicit, Consensus, and Semantic Scholar as specialized academic AI systems. Students work with all five and develop explicit comparative understanding of their different reliability and capability profiles. This comparative meta-literacy transfers as the tool landscape evolves in ways that proficiency with any single platform does not.

\subsection{The Dual-Enrollment Design}

The simultaneous enrollment of undergraduates (395) and graduate students (495) was a deliberate pedagogical choice rather than an administrative convenience. The heterogeneous classroom generates comparative learning that a homogeneous section cannot. Graduate students' domain expertise across a range of disciplines provides concrete, varied examples of how AI tools perform differently in different research contexts, observations that become shared instructional content through structured discussion. Undergraduates, meanwhile, benefit from early exposure to research competencies that are usually left to graduate training, and from witnessing how those competencies operate in advanced research practice. The differentiated assessment structure, described in Section~\ref{sec:assessment}, ensures that the shared instructional environment maintains appropriate challenge for both populations without compromising the integrity of either assessment standard.

\section{Module-Level Instructional Design}
\label{sec:modules}

The following sections describe each module's instructional logic,
central design decisions, and representative activities. The goal is not to reproduce a syllabus but to articulate the principled rationaleconnecting design choices to learning objectives within each module.

\subsection{Module A: AI-Assisted Comprehension of Individual Papers}

\subsubsection*{Design Logic}

Module A addresses the foundational layer of literature review competency: understanding individual papers with sufficient depth to evaluate their quality and relevance. This is the layer at which students most immediately want AI assistance and where the risks of scaffold dependency are most consequential. A student who accepts AI summaries as substitutes for reading may produce apparently competent work while failing to develop the analytical skills that more advanced research tasks require. The module's instructional strategy is therefore to use AI to \textit{accelerate and deepen} comprehension rather than replace it, a distinction that turns out to require explicit instruction because it is not intuitive.

\subsubsection*{Foundational Principles: AI Roles, Limitations, and
Verification}

The module opens by establishing the course's foundational conceptual framework: four AI roles in research (cognitive scaffold, efficiency multiplier, conceptual bridge, and pattern detector) and, with equal emphasis, the characteristic limitations of each. Students learn from the first session that LLMs generate plausible text rather than retrieving verified facts, that hallucinated citations are indistinguishable from real ones without independent checking, and that knowledge cutoff dates mean even recent-seeming AI claims about the literature may be outdated. This early, unambiguous establishment of limitations is a deliberate pedagogical choice: the course treats verification not as an optional quality-control step but as a constitutive element of responsible AI use, and that framing must be established before students develop habitual reliance on unverified outputs.\\
Prompt engineering is introduced through five operational principles: specificity, context provision, role assignment, output specification, and iterative refinement, and practiced through structured comparison exercises in which students apply vague and precise prompts to the same research task and analyze the difference in output quality and verifiability. The four pillars of ethical AI use that govern conduct throughout the course are introduced in this opening phase: transparency, verification, attribution, and authorial ownership.

\subsubsection*{Structural Literacy and Strategic Reading}

A central pedagogical contribution of Module A is making the strategic reading process explicit in a form that integrates AI assistance without substituting for it. Students learn a six-step reading procedure: reading the abstract independently; examining tables and figures; reading introduction and conclusion; using AI to explain unfamiliar methods orterminology identified in the preceding steps; reading methods and results with AI explanations in hand; and synthesizing conclusions independently before reviewing any AI summary to assess agreement. The AI-augmented steps are explicitly bracketed within a larger process that begins and ends with independent engagement. Students who follow this procedure use AI to access papers that might otherwise be technically opaque, while retaining the active engagement with primary sources that produces genuine comprehension.\\
Advanced prompting techniques for paper analysis are introduced
progressively: multi-step prompts that request structured extraction with page-number anchors, comparative prompts analyzing multiple papers simultaneously, and Socratic prompts that use AI to generate critical questions rather than summaries, stimulating analytical engagement rather than passive reception.

\subsubsection*{Methodology Evaluation}

The module's most technically demanding element is training students to evaluate research methodology with sufficient rigor to assess evidence quality. Students learn to identify and distinguish common study designs,to apply foundational evaluation questions to any study, and to identify methodological limitations beyond those the authors themselves acknowledge. The AI-Augmented Critical Reading Framework structures this process in four stages: AI-assisted initial extraction of methodological features, independent researcher classification of study design and quality, comparative analysis identifying discrepancies between the two, and integrated evaluation that draws on both while privileging researcher
judgment. Discrepancies between AI characterization and independent
assessment are treated as investigative signals rather than resolved automatically in favor of either source.\\
The module's culminating activity requires students to produce a
structured reading log for three papers on their chosen research topic, demonstrating the full Module A competency set: verified citations, independent design classification, key findings with page-number anchors, author-stated and researcher-identified limitations, and complete AI attribution documentation. This assignment establishes the documentation standard that all subsequent work maintains.

\subsection{Module B: AI-Assisted Construction and Validation of
Knowledge Taxonomies}

\subsubsection*{Design Logic}

Module B marks the first major scaling transition: from understanding individual papers to organizing knowledge across a collection. This transition introduces the defining challenge of literature review at scale, which is the need for systematic organizational structures that prevent important patterns from being lost as the corpus grows. The central concept of the module is the \textit{knowledge taxonomy}: a hierarchical, multi-dimensional structure that classifies research within a domain according to meaningful categories and relationships.\\
The module's central pedagogical argument is that good taxonomies do not merely organize papers; they reveal conceptual structure, expose methodological distributions, and, most importantly, identify the empty cells that signal research gaps. The last function connects Module B directly to Module C and establishes the taxonomy not as a terminal product but as an analytical instrument.

\subsubsection*{Taxonomy Design Principles}

Three taxonomy types are introduced and compared: typological taxonomies classifying by kind (e.g., intervention types), dimensional taxonomies organizing research along multiple simultaneous axes (e.g., intervention type by delivery mode by population), and hierarchical taxonomies nesting subcategories within broader categories. Students learn that dimensional taxonomies, while more complex to construct, are generally more analytically powerful because they reveal interaction patterns between organizational dimensions that single-axis classifications conceal.\\
Pattern recognition across five levels (convergent, divergent,
methodological, temporal, and population patterns) constitutes the
analytic core of multi-paper synthesis in this module. A critical
pedagogical contribution is the distinction between genuine and spurious convergence, a distinction that AI tools frequently obscure. AI tools tend to group papers by surface similarity (keyword matching) rather than conceptual alignment, producing apparent convergence among papers measuring related but meaningfully distinct constructs. Students learn to detect spurious convergence by examining whether ostensibly convergent papers are operationalizing the same construct, using comparable populations, and drawing on independent rather than overlapping samples, a form of methodological literacy that requires understanding research design well enough to see past AI-generated organizational patterns.

\subsubsection*{Validation}

The module's most rigorous component is taxonomy validation, the process that transforms an AI-generated draft structure into a scholarly artifact reliable enough to support synthesis. Five validation criteria are established: comprehensiveness, distinctness, meaningfulness, balance, and utility. The primary quantitative validation method is inter-rater reliability using Cohen's $\kappa$ \cite{landis1977measurement}, applied through stratified random sampling of 25\% of the paper set, independent classification, computation and interpretation of $\kappa$ against established benchmarks, and structured discrepancy resolution. Requiring quantitative validation is a deliberate choice: it makes the quality of AI-assisted organizational work falsifiable, and it introduces students to a standard of rigor that applies across research methods contexts well beyond the course.\\
The module's culminating activity produces a validated, multi-dimensional taxonomy with complete documentation of construction decisions, refinement history, inter-rater reliability statistics, and AI attribution. This artifact serves as the primary input to Module C.

\subsection{Module C: AI-Assisted Identification of Research Gaps and
Frontier Mapping}

\subsubsection*{Design Logic}

Module C marks the second major conceptual transition: from consuming and organizing existing research to contributing to the research conversation. This transition is explicitly named and discussed as an epistemological shift. Students who complete Module B have a comprehensive, validated organizational picture of a literature. Module C teaches them to read that picture for what is absent, interpreting empty cells and sparse areas not as limitations of their search process but as signals of genuine scholarly opportunity, and to evaluate the importance of those absences with the judgment that distinguishes productive research questions from unimportant ones.

\subsubsection*{Gap Identification and Importance Assessment}

A six-type knowledge gap taxonomy provides systematic structure for
identification: evidence gaps, methodological gaps, population gaps, contextual gaps, theoretical gaps, and translational gaps. Each type is characterized by its identifying features and the forms of evidence required to establish it as a genuine absence rather than an artifact of incomplete searching. The gap taxonomy is paired with a five-dimension importance rubric (theoretical importance, practical importance, feasibility, novelty, and coherence with research trajectory) that students apply to evaluate each candidate gap.\\
Importance assessment is explicitly distinguished from gap identification, because the two are frequently conflated: many gaps are genuine but unimportant, and learning to apply this distinction is a threshold-crossing moment for students who have been conditioned to treat any unstudied question as inherently worthy of investigation. AI tools are particularly prone to asserting the importance of gaps without substantiating the claim, making this distinction a critical site for verification discipline.\\
Frontier mapping complements gap analysis by using temporal citation network analysis to identify where active research is occurring and where opportunities exist at the edges of established knowledge. Students learn to use Semantic Scholar's citation tools to conduct forward citation searches, identify papers that are currently defining research frontiers through rapid citation accumulation, and distinguish genuine frontiers from keyword-driven publication surges driven by topic fashion rather than scientific momentum.

\subsubsection*{Hallucination Typology and Verification Discipline}

Module C concludes with a systematic treatment of AI hallucinations and output biases, deliberately positioned at the transition between gap identification and synthesis writing. By this stage of the course, students have used AI tools extensively on their own research topics and have encountered hallucinations firsthand. The systematic typology of hallucination types encountered at this stage consequently functions not as abstract warning but as retrospective explanation, providing vocabulary and structure for phenomena students have already observed, a sequencing principle discussed further in Section~\ref{sec:lessons}.\\
Five hallucination types are examined: citation hallucinations,
statistical hallucinations, methodological hallucinations, theoretical hallucinations, and synthesis hallucinations, including false consensus and false gap claims. For each type, students learn the characteristic detection signals and the required verification procedures. Six sources of systematic AI output bias are addressed in parallel: recency bias, English-language bias, open-access bias, citation-count bias, framing bias, and confirmation bias in gap assessment. Students learn to treat each not as a potential problem but as a predictable systematic distortion requiring active correction.\\
The module's culminating activity produces a gap analysis report
documenting five prioritized gaps, each classified by type, evaluated on the importance rubric with justification, and accompanied by a frontier map showing temporal dynamics and a research question with novelty and feasibility assessment.

\subsection{Module D: Synthesis and Literature Review Production}

\subsubsection*{Design Logic}

Module D integrates all competencies developed in Modules A--C into the production of complete literature reviews. The module's central challenge is a distinction that AI tools actively obscure: the difference between AI-assisted and AI-authored writing. At the synthesis stage, AI can generate prose that sounds like scholarly synthesis while being substantively inadequate: organized by keywords rather than arguments, listing findings rather than integrating them, and lacking the interpretive perspective that communicates a scholar's own analytical position.\\
The module frames authorial voice not as a stylistic preference but as a substantive epistemological competency. A literature review is a structured argument, not a catalog, and the argument must be the researcher's own. AI can help construct the scaffold; only the researcher can supply the argument.

\subsubsection*{Synthesis Structure and GenAI Roles}

Five organizational structures for literature reviews are introduced and compared: chronological, methodological, thematic, theoretical, and problem-solution. Students learn not only the features of each structure but the selection criteria that determine which structure is appropriate for a given body of literature and research question, a form of judgment that AI tools cannot exercise reliably because it requires understanding the literature's conceptual topology.\\
Synthesis techniques are introduced on a spectrum from weakest to strongest: paper-by-paper summary, vote counting, narrative synthesis, and tabular and visual synthesis. Five legitimate GenAI roles at the synthesis stage are identified: structured data extraction into evidence tables, pattern detection across study features, draft paragraph generation for researcher revision, organizational structure suggestion, and internal consistency checking. Each of these roles is explicitly bounded: AI-generated prose that enters the review without substantial researcher revision is not the researcher's work.\\
Two practical tests for appropriate AI use in synthesis are introduced: the revision depth test (has the researcher evaluated every factual claim in the AI draft?) and the deletion test (could the researcher reconstruct the passage without the AI draft?). Students apply these tests in peer review activities and use them to evaluate their own writing process.

\subsubsection*{Attribution and Professional Standards}

Module D addresses AI attribution requirements for final manuscripts, introducing students to emerging journal disclosure standards and providing a standardized attribution template that can be adapted for journal submission. The methods section of the literature review, which requires documentation of which tools were used, for which tasks, and with what verification procedures and error correction, is written without AI assistance, a requirement discussed in Section~\ref{sec:lessons} that proved unexpectedly diagnostic of students' understanding of their own research process.\\
The module concludes with final presentations to the class, through which students develop the oral communication skills associated with scholarly presentation and receive peer feedback across disciplinary boundaries, an experience that reinforces the course's discipline-agnostic premise by demonstrating that research communication competencies transfer across
fields.

\section{Assessment Architecture and Ethics Infrastructure}
\label{sec:assessment}

\subsection{Philosophy and Grade Distribution}

The course grade comprises four components: Pre- and Post-Course
Assessment Surveys (5\%), Weekly Assignments (60\%), Peer Evaluations and Participation (10\%), and Final Literature Review (25\%). This distribution operationalizes the course's central pedagogical commitment: in research methods education, process matters as much as product. Students who complete high-quality weekly assignments with documented, verified, attributed AI use are developing the skills that produce quality final work; weighting process-level assessment heavily incentivizes consistent engagement rather than end-of-semester consolidation.

\subsection{Weekly Assignments and the AI Use Log}

Weekly assignments are the primary vehicle for formative assessment and skill development. Each assignment requires students to apply the current week's skills to their chosen research topic and to submit complete AI use documentation alongside the substantive work. Documentation requirements include exact prompts, AI-generated outputs, verification steps, and discrepancies found. Grading addresses both substantive quality and documentation completeness, operationalizing the principle that undocumented AI use constitutes an integrity concern independent of the quality of the output it produced.\\
The AI use log, maintained throughout the semester and submitted with the final project, creates a longitudinal record of each student's tool use, prompt development, and verification practice. Analysis of these logs across the cohort constitutes the primary source of the implementation observations reported in Section~\ref{sec:lessons}.

\subsection{Assessment Rubrics for AI-Assisted Work}

Assessing AI-assisted work requires rubrics that evaluate process quality as well as product quality. The course employs a four-dimension rubric for all major assignments: \textit{substantive quality} (accuracy, completeness, and analytical rigor of the submitted work); \textit{verification quality} (whether factual claims, citations, and statistics have been independently confirmed); \textit{attribution completeness} (whether AI contributions are fully documented with the specificity the course template requires); and \textit{authorial mastery} (whether the student can explain, justify, and defend every element of their submission).\\
The authorial mastery dimension is assessed primarily through
participation and peer review: a student who cannot explain how their work was produced or defend its claims under critical questioning has not met the standard, regardless of the surface quality of the submitted text. This dimension directly operationalizes the verified engagement principle and the threshold concept of AI as scaffold rather than substitute.

\subsection{The Final Literature Review}

The culminating assessment is an original, independently authored
literature review produced through the full four-module process.
Undergraduate (395) students produce reviews of approximately
3,000--4,000 words surveying a defined research area; graduate (495) students produce 4,000--6,000 word reviews with more extensive gap analysis and a developed future research agenda. All reviews include a complete AI attribution statement in the methods section and an acknowledgments disclosure.\\
The final review is assessed on the four-dimension rubric, with the
substantive quality and authorial mastery dimensions weighted more
heavily for the final product than for weekly assignments. The AI
attribution statement is evaluated separately against a higher standard, reflecting its function as a professional document with direct applicability to journal submission.

\subsection{Ethics as Subject, Not Policy}

A defining design decision was to treat AI ethics and integrity not as a course policy (rules about what is and is not permitted) but as a course \textit{subject}: a domain of knowledge and judgment that instruction explicitly develops. Most courses address AI by restricting it through policy. BSTA~495/395 addresses it by teaching students to analyze the ethical dimensions of their own AI practices and to develop the principled judgment to navigate those dimensions independently, without recourse to rules.\\
This approach is grounded in the observation that policy compliance is a weaker basis for integrity than principled understanding. A student who meets attribution requirements because they understand how proper attribution protects the infrastructure of cumulative scholarly knowledge is more reliably ethical (and more durably so) than one who complies to avoid penalty. The former carries their understanding across contexts; the latter carries only the specific rules they learned, applied only where enforcement is visible.

\subsection{The Standardized AI Attribution Template}

The course's standardized AI attribution template operationalizes
transparency requirements in a form students can adopt beyond the
course. For each AI interaction, the template records the tool and version, the task purpose, the exact prompt used, the output received, the verification method applied, the error rate found, and how errors were corrected. Completed templates are submitted with every assignment.\\
The template serves pedagogical functions beyond compliance. It creates a structured reflection practice, requiring students to articulate what they asked AI to do and whether it succeeded. It documents the development of prompt engineering competency across the semester: early prompts tend to be vague and single-sentence; late prompts incorporate role assignment, context provision, and structured output specification, providing a concrete developmental record. And it gives students a model for the AI disclosure statements that a growing number of journals now require, a directly applicable professional skill.

\subsection{The Responsible Reliance Spectrum}

Each module introduces a responsible reliance spectrum mapping common AI use patterns onto a scale from minimal-concern uses (explaining unfamiliar terminology, generating initial search terms) through moderate-concern uses (taxonomy category generation, abstract extraction) to high-concern uses (synthesis paragraph drafting without verification, novelty assessment without independent literature checking). Students learn to recognize which spectrum position their current AI use occupies and to apply verification effort proportional to the concern level. This spectrum provides a decision-making tool that operates independently of specific tools or tasks, supporting the transfer of verification judgment to novel AI use situations students will encounter throughout their research careers.

\section{Implementation Observations}
\label{sec:lessons}

The following observations draw on AI use log analysis, assignment
performance, and instructor reflection from the Spring 2026 inaugural offering. They are reported not as outcome data (formal analysis is ongoing) but as design-relevant findings that bear on the validity of the course's instructional principles and that informed post-hoc refinements for future offerings.

\subsection{The Experience-Before-Taxonomy Sequencing Effect}

The placement of the systematic hallucination and bias module at the transition between Modules C and D (rather than early in the course as a precautionary introduction) proved to be one of the most pedagogically consequential design decisions. By the time students reached this content, they had used AI tools for several weeks on their own research topics and had encountered hallucinations firsthand: citations that did not exist, statistics not present in the papers described, patterns asserted with confidence that careful reading contradicted. The systematic typology of hallucination types consequently functioned as retrospective explanation rather than prospective warning, providing vocabulary and structure for phenomena students had already observed and been puzzled by.\\
This experience-before-taxonomy sequencing produced deeper engagement with the verification concepts than earlier placement would have. Students engaged with hallucination types not as abstract categories but as descriptions of their own empirical experiences, and the associated verification procedures carried the weight of practical necessity rather than cautionary recommendation. Future curriculum designers working in this space may find this a useful general principle: AI failure modes are better taught after students have accumulated enough tool experience to recognize what they are being taught.

\subsection{Cross-Disciplinary Learning Dynamics}

The decision to enroll students across disciplines produced learning dynamics that a discipline-specific section could not have generated. Students discovered through shared discussion that AI hallucination rates and failure modes differed by research domain: citation hallucinations were more frequent and more consequential in narrow specialty areas where LLM training data was sparse, while pattern identification operated differently in quantitative and qualitative research literatures. The six gap types had different relative prevalence across fields, and the distinction between genuine and spurious convergence required different evidential standards in different disciplinary contexts.\\
These observations were not planned instructional content; they emerged organically from cross-disciplinary discussion and were subsequently integrated into the course's treatment of AI bias and verification. The heterogeneous classroom functioned as a generative source of comparative evidence that enriched the course's own analytical content.

\subsection{Prompt Engineering Development Across the Semester}

Analysis of AI use logs across the cohort revealed consistent
improvement in prompt sophistication from early to late assignments. Early prompts characteristically took the vague, single-sentence form (``Summarize this paper''; ``What are the main themes?''). Late prompts incorporated role assignment, context provision, structured output specification, and iterative dialogue in ways that substantially improved both output quality and the ease of verification. This development was associated with reductions in the verification errors students reported encountering, a pattern consistent with the interpretation that better prompts produce more constrained, more checkable outputs.\\
The implication for curriculum design is that prompt engineering
content benefits from distributed presentation across the course
rather than concentrated early introduction. The course revisits and extends prompt engineering technique at each module transition,
aligning instruction with the increasing complexity of the AI tasks
students are performing. The initial session establishes the
framework; subsequent modules introduce advanced techniques
(chain-of-thought prompting, structured extraction formats, and
multi-step dialogue) as students encounter tasks that require them.

\subsection{The Methods Section as a Diagnostic}

Requiring students to write the methods sections of their final
literature reviews without AI-generated text proved unexpectedly
diagnostic of the quality of their research process understanding.
Students who had engaged substantively with their own process, having made deliberate decisions about which tools to use for which tasks and documented those decisions in their AI use logs, produced accurate, specific methods sections that read as genuine methodological accounts. Students who had relied on AI for process management, rather than using AI as a tool within a researcher-managed process, produced methods descriptions that were generic, procedurally vague, or inaccurate about the verification steps actually taken.\\
This observation has implications beyond the course. The ability to
give an accurate account of one's own research process is a
professional competency that the field increasingly requires (journal AI disclosure requirements are moving in the direction of greater specificity), and it is an ability that is not developed by AI use without reflection. The methods writing requirement operationalizes verified engagement at the level of research process rather than individual claim.

\section{Preliminary Outcome Evidence}
\label{sec:outcomes}

\subsection{Overview and Survey Instrument}

This section reports preliminary outcome evidence from the Spring 2026 inaugural offering of BSTA~495/395, drawn from pre- and post-course self-report surveys administered in Weeks~1 and~13, respectively. The pre-course instrument assessed eight confidence dimensions directly aligned with the course's four-module learning objectives, current AI tool use behavior, prior literature review experience, and familiarity with AI ethics guidelines. The post-course instrument repeated the eight confidence items, added six learning outcome agreement items operationalizing the course's competency framework, and included two open-response items soliciting student-identified learning gains and improvement suggestions.\\
Because this is a single-cohort, pre--post design without a control condition, all findings are presented as preliminary indicators of student-perceived learning gains consistent with the course's intended outcomes, not as causal evidence of instructional effectiveness. Formal inferential analysis and qualitative theme development are ongoing and will be reported in a dedicated outcomes paper.

\subsection{Sample Characteristics}

\begin{figure}[htbp]
  \centering
  \includegraphics[width=\textwidth]{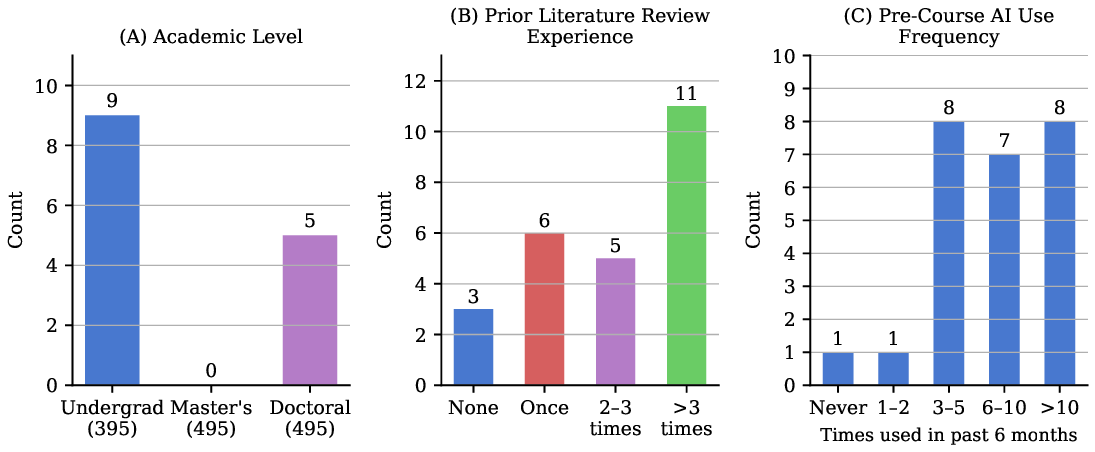}
  \caption{Sample background characteristics ($n = 27$, pre-survey).
  Panel~(A) shows enrollment by academic level, distinguishing
  traditional undergraduates, accelerated 4+1 students registered
  at the undergraduate level, and doctoral students; Panel~(B) shows
  prior literature review experience; Panel~(C) shows pre-course AI
  tool use frequency in the preceding six months.}
  \label{fig:background}
\end{figure}

Twenty-seven students completed the pre-course survey ($n = 27$).
A small number of students withdrew from the course after Week~1,
following pre-survey administration, and a small number joined after the semester began, resulting in a partially overlapping post-course sample ($n = 29$). Wilcoxon signed-rank tests were computed on matched pairs only, using the subset of students who completed both surveys ($n = 26$); unmatched respondents contributed to descriptive post-course statistics but were excluded from all paired comparisons. The three additional post-survey respondents were students who enrolled after the pre-survey administration date but who completed the full 13-week course; they did not differ from the original cohort in academic level or prior research experience. Their exclusion from matched analyses is conservative and does not affect the interpretation of any paired result, but their inclusion in descriptive post-course means should be noted when interpreting the magnitude of post course distributions. Figure~\ref{fig:background} presents the sample's background characteristics. The cohort comprised nine traditional undergraduates (33\%), thirteen students enrolled in the accelerated 4+1 undergraduate-to-Master's program (48\%) who registered at the undergraduate level (BSTA~395), and five doctoral students (19\%), consistent with the dual-enrollment design. Table~\ref{tab:sample} provides complete sample descriptives.\\
The majority of students (89\%) entered the course having used generative AI tools at least three times in the preceding six months, with ChatGPT the dominant prior tool. However, the academic AI platforms central to the course (Elicit, Consensus, Semantic Scholar) were largely unfamiliar to enrolled students, confirming that the course was addressing a genuine competency gap rather than providing redundant instruction. Notably, 41\% of entering students reported having conducted more than three literature reviews previously, suggesting substantial prior research experience in a majority of the cohort. Despite this experience, 82\% rated themselves only slightly or moderately familiar with AI ethics guidelines for academic research, establishing the course's ethics infrastructure as non-redundant for experienced as well as novice researchers. The primary stated course goal for 70\% of enrollees was learning to use AI tools responsibly.

\FloatBarrier

\begin{table}[htbp]
\centering
\begin{threeparttable}
\caption{Sample Characteristics.}
\label{tab:sample}
\begin{tabular}{p{9cm}ccc}
\toprule
\textbf{Characteristic} & \textbf{\textit{n}} & \textbf{\%} &
\textbf{\% of Post} \\
\midrule
\multicolumn{4}{l}{\textit{Enrollment \& Completion}} \\
\quad Total enrolled        & 27 & \multicolumn{1}{c}{---} &
  \multicolumn{1}{c}{---} \\
\quad Completed pre-survey  & 27 & \multicolumn{1}{c}{---} & 100 \\
\quad Completed post-survey & 29 & \multicolumn{1}{c}{---} & 107 \\[4pt]
\multicolumn{4}{l}{\textit{Academic Level}} \\
\quad Undergraduate -- standard (395)        &  9 & 33 &
  \multicolumn{1}{c}{---} \\
\quad Undergraduate -- accelerated 4+1 (395) & 13 & 48 &
  \multicolumn{1}{c}{---} \\
\quad Doctoral student (495)                 &  5 & 19 &
  \multicolumn{1}{c}{---} \\[4pt]
\multicolumn{4}{l}{\textit{Prior Literature Review Experience}} \\
\quad None              &  3 & 11 & \multicolumn{1}{c}{---} \\
\quad Once              &  6 & 22 & \multicolumn{1}{c}{---} \\
\quad 2--3 times        &  5 & 19 & \multicolumn{1}{c}{---} \\
\quad More than 3 times & 11 & 41 & \multicolumn{1}{c}{---} \\[4pt]
\multicolumn{4}{l}{\textit{Pre-Course AI Use Frequency (past 6 months)}} \\
\quad Never              &  1 &  4 & \multicolumn{1}{c}{---} \\
\quad 1--2 times         &  1 &  4 & \multicolumn{1}{c}{---} \\
\quad 3--5 times         &  8 & 30 & \multicolumn{1}{c}{---} \\
\quad 6--10 times        &  7 & 26 & \multicolumn{1}{c}{---} \\
\quad More than 10 times &  8 & 30 & \multicolumn{1}{c}{---} \\[4pt]
\multicolumn{4}{l}{\textit{Familiarity with AI Ethics Guidelines}} \\
\quad Not at all familiar &  2 &  7 & \multicolumn{1}{c}{---} \\
\quad Slightly familiar   & 11 & 41 & \multicolumn{1}{c}{---} \\
\quad Moderately familiar & 11 & 41 & \multicolumn{1}{c}{---} \\
\quad Very familiar       &  1 &  4 & \multicolumn{1}{c}{---} \\[4pt]
\multicolumn{4}{l}{\textit{Primary Course Goal}} \\
\quad Responsible AI use         & 19 & 70 & \multicolumn{1}{c}{---} \\
\quad Improve research skills    &  3 & 11 & \multicolumn{1}{c}{---} \\
\quad Literature review / thesis &  2 &  7 & \multicolumn{1}{c}{---} \\
\quad Other                      &  1 &  4 & \multicolumn{1}{c}{---} \\
\bottomrule
\end{tabular}
\begin{tablenotes}
\small
\item \textit{Note.} $n = 27$ for pre-survey; $n = 29$ for
post-survey. \% of Post\,=\,percentage of post-survey respondents
relative to pre-survey enrollment; ---\,=\,not applicable. Academic
levels reflect the dual-enrollment design (395\,=\,undergraduate;
495\,=\,graduate). Students in the accelerated 4+1 undergraduate-to-Master's program registered under the undergraduate section (BSTA~395) and are counted accordingly; no students held traditional Master's status.
\end{tablenotes}
\end{threeparttable}
\end{table}
\FloatBarrier

\subsection{Pre--Post Confidence Gains}

\begin{figure}[htbp]
  \centering
  \includegraphics[width=\textwidth]{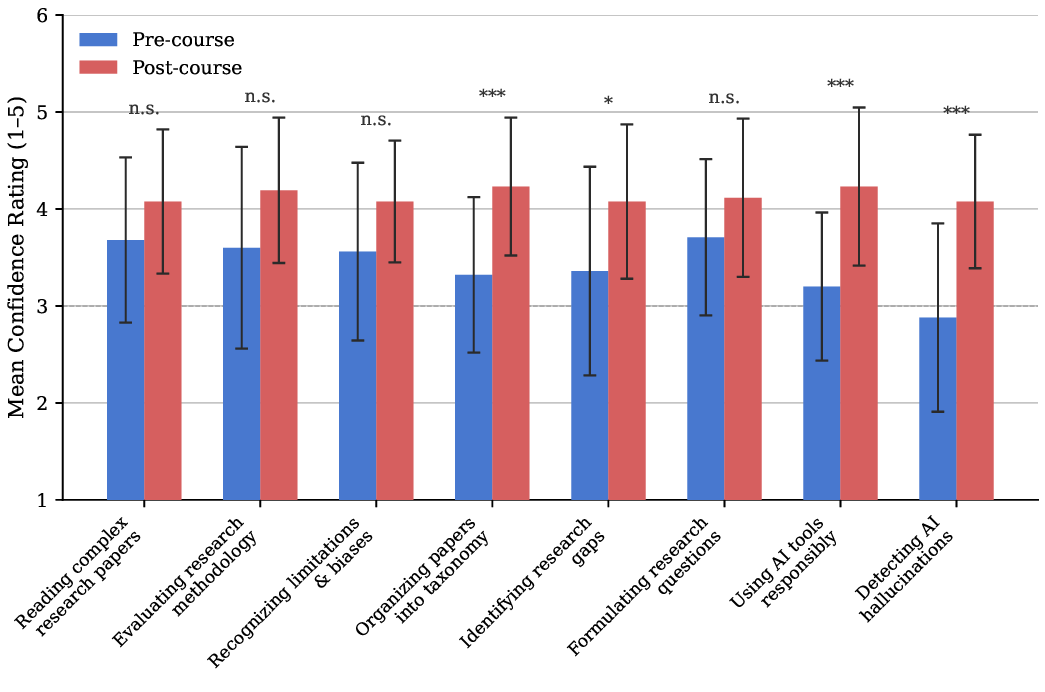}
  \caption{Pre- and post-course mean confidence ratings by competency domain. Error bars represent one standard deviation. Significance markers derive from Wilcoxon signed-rank tests on matched pairs ($n = 26$); unmatched post-survey respondents are included in post-course means but excluded from significance testing: $^{*}p < .05$, $^{**}p < .01$, $^{***}p < .001$,
  n.s.\,$p \geq .05$.}
  \label{fig:confidence}
\end{figure}

\begin{figure}[htbp]
  \centering
  \includegraphics[width=\textwidth]{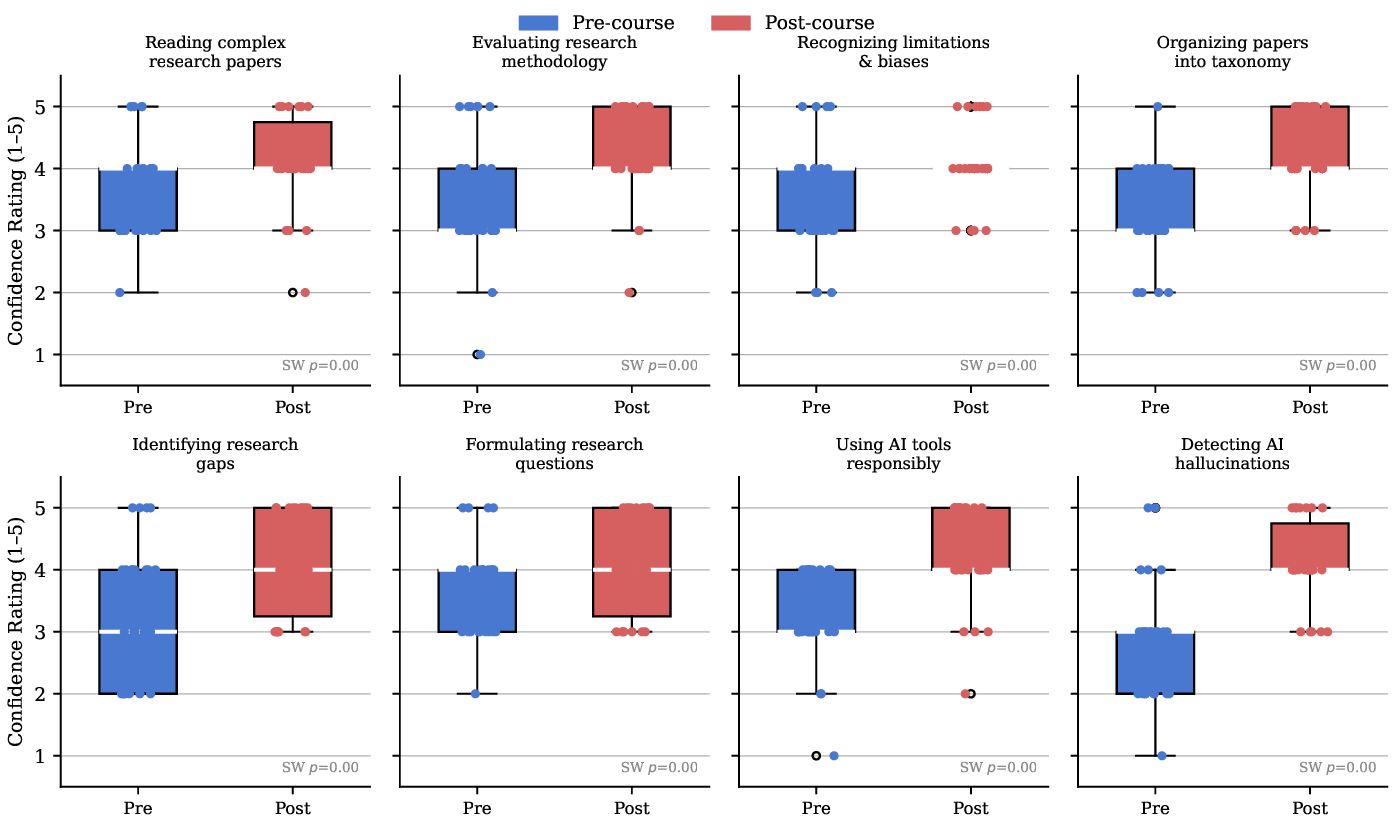}
  \caption{Distribution of confidence ratings before and after the
  course for all eight competency domains. Boxes indicate the
  interquartile range; whiskers extend to 1.5 times the IQR;
  individual data points are overlaid with jitter. SW\,=\,Shapiro--Wilk normality test on post-course values; all post-course distributions departed significantly from normality, motivating the use of Wilcoxon signed-rank tests for significance testing.}
  \label{fig:boxplots}
\end{figure}

Across the eight confidence items, mean confidence increased from
$M_{\mathrm{pre}} = 3.41$ ($SD = 0.93$) to $M_{\mathrm{post}} = 4.13$ ($SD = 0.74$), a grand mean gain of $+0.72$ points on the five-point scale. Figure~\ref{fig:confidence} presents the item-level pre- and post-course means with error bars and significance annotations. Figure~\ref{fig:boxplots} shows the full distribution of responses for each item before and after the course. Table~\ref{tab:confidence} presents complete descriptive statistics, effect sizes, and Wilcoxon signed-rank $p$-values for all ten confidence and skills items.

\FloatBarrier

\FloatBarrier

\begin{table}[htbp]
\centering
\caption{Pre- and Post-Course Confidence and Skills Ratings}
\label{tab:confidence}
\begin{threeparttable}
\resizebox{0.95\textwidth}{!}{%
\begin{tabular}{p{6cm}ccccc}
\toprule
\textbf{Competency Domain}
  & \textbf{Pre} $M$ (SD)
  & \textbf{Post} $M$ (SD)
  & $\Delta M$
  & \textbf{Cohen's} $d$
  & $p$ \\
\midrule
\multicolumn{6}{l}{\textit{Confidence Items (Q1--8)}} \\
Reading complex research papers
& 3.68 (0.84) & 4.08 (0.73) & +0.40 & +0.51 & 0.166 \\
Evaluating research methodology
& 3.60 (1.02) & 4.19 (0.73) & +0.59 & +0.67 & 0.063 \\
Recognizing limitations and biases
& 3.56 (0.90) & 4.08 (0.62) & +0.52 & +0.67 & 0.059 \\
Organizing papers into a taxonomy
& 3.32 (0.79) & 4.23 (0.70) & +0.91 & +1.23 & $< 0.001$ \\
Identifying research gaps
& 3.36 (1.05) & 4.08 (0.78) & +0.72 & +0.77 & 0.028 \\
Formulating research questions
& 3.71 (0.79) & 4.12 (0.80) & +0.41 & +0.51 & 0.084 \\
Using AI tools responsibly
& 3.20 (0.75) & 4.23 (0.80) & +1.03 & +1.33 & $< 0.001$ \\
Detecting AI hallucinations
& 2.88 (0.95) & 4.08 (0.67) & +1.20 & +1.45 & $< 0.001$ \\
\multicolumn{6}{l}{\textit{Skills Assessment Items
  (Q13--14 pre / Q9--10 post)}} \\
Verifying AI outputs
& 3.08 (0.89) & 4.08 (0.73) & +1.00 & +1.22 & $< 0.001$ \\
AI attribution practice
& 2.12 (0.82) & 4.12 (0.85) & +2.00 & +2.40 & $< 0.001$ \\
\bottomrule
\end{tabular}%
}
\begin{tablenotes}
\footnotesize
\item
\begin{minipage}{0.95\textwidth}
\textit{Note.} Scale: 1\,=\,not at all confident to 5\,=\,extremely confident. $\Delta M$\,=\,post minus pre mean. Cohen's $d$ computed using pooled SD. $p$-values from Wilcoxon
signed-rank tests on matched pairs ($n = 26$); students who
completed only one survey contributed to descriptive means but
were excluded from paired significance tests. Pre: $n = 27$; Post (descriptive): $n = 26$--29.
\end{minipage}
\end{tablenotes}
\end{threeparttable}
\end{table}
\FloatBarrier

\begin{figure}[htbp]
  \centering
  \includegraphics[width=0.80\textwidth]{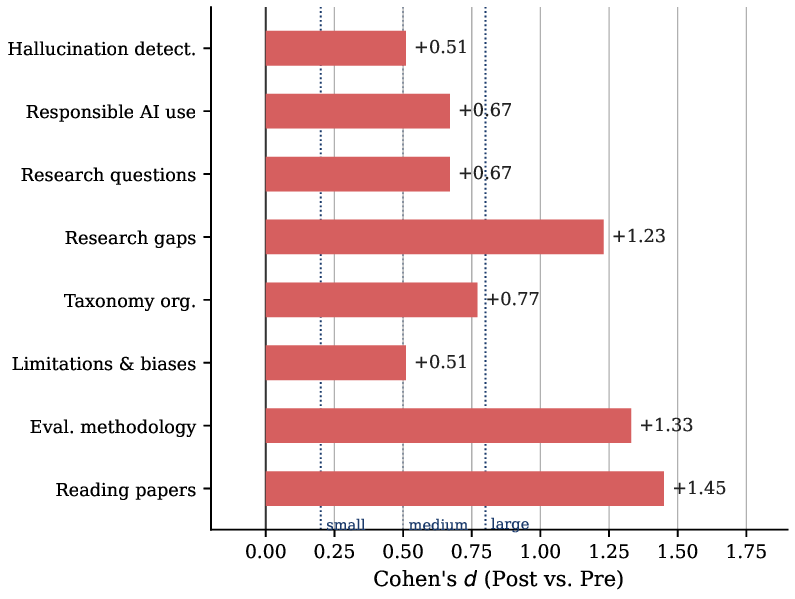}
  \caption{Effect sizes (Cohen's $d$) for pre--post confidence gains across the eight competency domains. Dashed vertical lines indicate conventional thresholds for small ($d = 0.2$), medium ($d = 0.5$), and large ($d = 0.8$) effects \cite{cohen1988statistical}.}
  \label{fig:effectsizes}
\end{figure}

The three largest and most statistically robust gains are theoretically significant in light of the course's design. Hallucination detection showed the largest gain among the eight confidence items ($\Delta M = +1.20$, $d = +1.45$, $p < .001$), a very large effect reflecting the course's sustained, module-spanning emphasis on recognizing and verifying AI failure modes. Responsible AI use showed the second largest gain ($\Delta M = +1.03$, $d = +1.33$, $p < .001$), consistent with the course's ethics-as-subject infrastructure and the transparent documentation requirements embedded in every assignment. Taxonomy organization gained substantially ($\Delta M = +0.91$, $d = +1.23$, $p < .001$), directly validating Module B's design. Research gap identification showed a large and statistically significant gain ($\Delta M = +0.72$, $d = +0.77$, $p = .028$), consistent with Module C's focus on this competency. Figure~\ref{fig:effectsizes} presents a forest plot of effect sizes across all eight items.\\
Three items showed moderate gains that did not reach conventional
significance thresholds: reading research papers ($d = +0.51$,
$p = .166$), evaluating methodology ($d = +0.67$, $p = .063$), and
recognizing limitations and biases ($d = +0.67$, $p = .059$). These
results warrant interpretive care. Students entering the course with already moderate mean confidence in these domains (pre-course means of 3.68, 3.60, and 3.56, respectively) may have less room for
self-reported gain, a ceiling-approach effect consistent with the
distributions visible in Figure~\ref{fig:boxplots}. These two interpretations are in tension and bear distinguishing. The pre-course means on these three items (3.68, 3.60, and 3.56, respectively) were the three highest in the set, and all exceeded the grand pre-course mean of 3.41, consistent with a ceiling-approach effect in which students entering with above-average confidence have less room for self-reported gain. This distributional evidence favours the ceiling-approach interpretation over the epistemic-humility alternative, though the two are not mutually exclusive: some attenuation of post-course confidence is consistent with Module A's emphasis on the genuine difficulty of rigorous methodology evaluation, and a small downward recalibration among students who entered overconfident would be theoretically expected and pedagogically desirable. The distinction matters for replication: institutions implementing comparable curricula should expect smaller self-reported gains on competency domains where entering students report relatively high prior confidence, and should not interpret such results as instructional failure.

\FloatBarrier

\subsection{Skills Assessment Items}

\begin{figure}[htbp]
  \centering
  \includegraphics[width=0.75\textwidth]{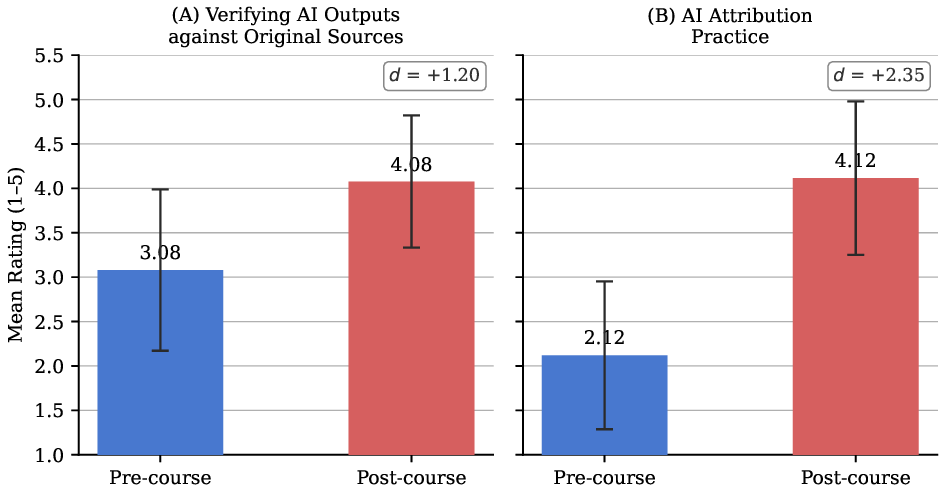}
  \caption{Pre- and post-course mean ratings for the two skills
  assessment items. Error bars represent one standard deviation.
  Cohen's $d$ values are shown in the upper right of each panel.}
  \label{fig:skills}
\end{figure}

The two skills assessment items, measuring practical ability to verify AI outputs and to attribute AI assistance correctly, showed the largest gains in the entire dataset. AI attribution practice rose from $M = 2.12$ ($SD = 0.82$) pre-course to $M = 4.12$ ($SD = 0.85$) post-course, a $\Delta M = +2.00$ gain with a very large effect size ($d = +2.40$, $p < .001$). This near-floor-to-ceiling shift represents the single most pronounced change across all measured competencies and is a direct validation of the course's ethics-as-subject infrastructure and the standardized AI attribution template embedded in every assignment. Verification of AI outputs similarly showed a very large gain ($\Delta M = +1.00$, $d = +1.22$, $p < .001$). Figure~\ref{fig:skills} presents the pre--post comparison for both skills items. Table~\ref{tab:effectsizes} provides the complete effect size summary across all ten items.

\FloatBarrier

\begin{table}[htbp]
\centering
\caption{Effect Size Summary for All Pre--Post Comparisons}
\label{tab:effectsizes}
\resizebox{0.95\textwidth}{!}{%
\begin{tabular}{p{6.5cm}cccc}
\hline
\textbf{Competency Item}
  & $\Delta M$
  & $d$
  & \textbf{Classification}
  & \textbf{Sig.} \\
\hline
Reading complex research papers
& +0.40 & +0.51 & Medium     & n.s. \\
Evaluating research methodology
& +0.59 & +0.67 & Medium     & n.s. \\
Recognizing limitations and biases
& +0.52 & +0.67 & Medium     & n.s. \\
Organizing papers into a taxonomy
& +0.91 & +1.23 & Very large & *** \\
Identifying research gaps
& +0.72 & +0.77 & Medium     & * \\
Formulating research questions
& +0.41 & +0.51 & Medium     & n.s. \\
Using AI tools responsibly
& +1.03 & +1.33 & Very large & *** \\
Detecting AI hallucinations
& +1.20 & +1.45 & Very large & *** \\
Verifying AI outputs
& +1.00 & +1.22 & Very large & *** \\
AI attribution practice
& +2.00 & +2.40 & Very large & *** \\
\hline
\end{tabular}%
}
\vspace{4pt}
\footnotesize
\textit{Note.} Cohen's $d$ classification: $< 0.2$ negligible,
0.20--0.49 small, 0.50--0.79 medium, 0.80--1.19 large,
$\geq 1.20$ very large. Significance from Wilcoxon signed-rank test. *** $p < .001$, ** $p < .01$, * $p < .05$, n.s.\ $p \geq .05$.
\end{table}
\FloatBarrier

\subsection{Post-Course Learning Outcome Ratings}

\begin{figure}[htbp]
  \centering
  \includegraphics[width=\textwidth]{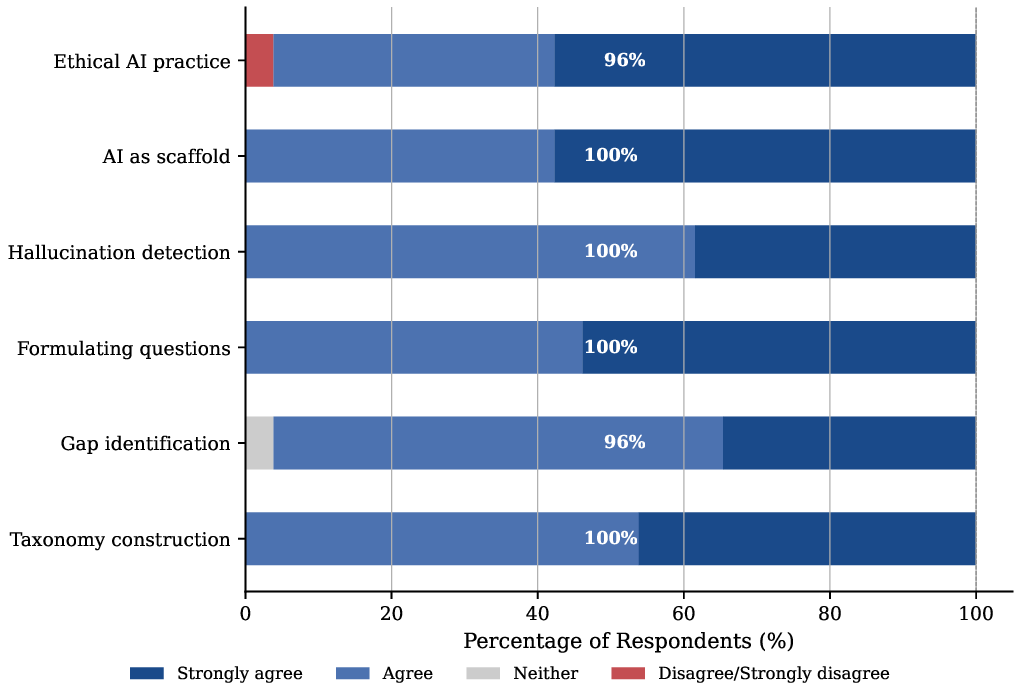}
  \caption{Post-course learning outcome ratings ($n = 26$). Bars show the percentage of respondents endorsing each response category. White bold labels inside bars indicate the combined
  agree-plus-strongly-agree percentage for each item.}
  \label{fig:outcomes}
\end{figure}

\begin{figure}[htbp]
  \centering
  \includegraphics[width=0.80\textwidth]{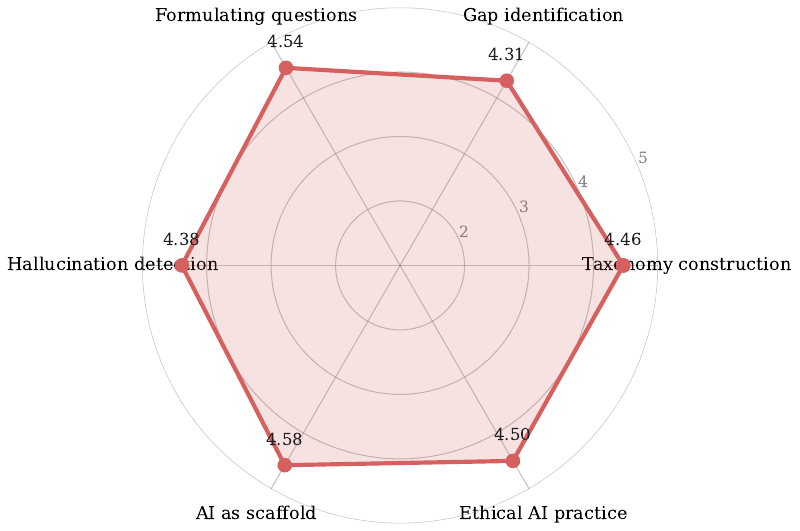}
  \caption{Post-course competency mastery profile ($n = 26$). Each
  vertex represents the mean agreement rating (1--5 scale) for one
  of the six learning outcome items. The shaded region illustrates
  the overall competency profile.}
  \label{fig:radar}
\end{figure}

Post-course learning outcome items assessed perceived mastery of six competencies directly operationalizing the course's four-module
framework on a five-point agreement scale. Results were uniformly
strong. Figure~\ref{fig:outcomes} presents the stacked distribution
of responses across all six items, and Figure~\ref{fig:radar} presents the mean mastery profile in radar format. Table~\ref{tab:outcomes} provides complete descriptive statistics and agree/strongly-agree percentages.

\FloatBarrier

\FloatBarrier

\begin{table}[htbp]
\centering
\caption{Post-Course Learning Outcome and Course Impact Ratings}
\label{tab:outcomes}
\resizebox{\textwidth}{!}{%
\begin{tabular}{p{6.5cm}cccc}
\hline
\textbf{Item}
  & \textbf{$M$ (SD)}
  & \textbf{Agree/SA \%}
  & \textbf{SA only \%}
  & \textbf{Primary Module} \\
\hline
\multicolumn{5}{l}{\textit{Learning Outcome Items}} \\
Taxonomy construction (Q13)
& 4.46 (0.51) & 100\% & 46\% & Module B \\
Gap identification (Q14)
& 4.31 (0.55) & 96\%  & 35\% & Module C \\
Formulating questions (Q15)
& 4.54 (0.51) & 100\% & 54\% & Module C \\
Hallucination detection (Q16)
& 4.38 (0.50) & 100\% & 38\% & Module C \\
AI as scaffold (Q17)
& 4.58 (0.50) & 100\% & 58\% & All modules \\
Ethical AI practice (Q18)
& 4.50 (0.71) & 96\%  & 58\% & All modules \\
\multicolumn{5}{l}{\textit{Course Impact Indicators}} \\
Overall course value (Q19)
& 3.92 (0.80) & 73\%  & 23\% & --- \\
Likelihood of applying skills (Q20)
& 4.31 (0.68) & 88\%  & 42\% & --- \\
AI tool usefulness (Q12)
& 4.46 (0.58) & 96\%  & 50\% & --- \\
\hline
\end{tabular}%
}
\vspace{4pt}
\footnotesize
\textit{Note.} Learning outcome items: scale 1 = strongly disagree to 5 = strongly agree ($n = 26$). Course impact items: value scale 1 = not valuable to 5 = extremely valuable; likelihood scale 1 = very unlikely to 5 = very likely; usefulness scale 1 = not useful to 5 = extremely useful. SA = strongly agree/extremely valuable/very likely.
\end{table}
\FloatBarrier

Four of the six competency items produced 100\% agree-or-strongly-agree endorsement: taxonomy construction ($M = 4.46$, $SD = 0.51$),
formulating research questions ($M = 4.54$, $SD = 0.51$), hallucination detection ($M = 4.38$, $SD = 0.50$), and AI as cognitive scaffold ($M = 4.58$, $SD = 0.50$). Gap identification and ethical AI practice each produced 96\% agree-or-strongly-agree rates ($M = 4.31$ and $M = 4.50$, respectively), with a single student selecting ``neither'' on each. The near-ceiling distributions across all six items suggest that students perceived the course as effectively developing the competencies it targeted. The relatively higher mean for AI as scaffold ($M = 4.58$) is consistent with the course's explicit, module-spanning emphasis on the threshold concept distinguishing AI-assisted from AI-replaced cognition.

\subsection{Course Impact and Future Application}

\begin{figure}[htbp]
  \centering
  \includegraphics[width=0.8\textwidth]{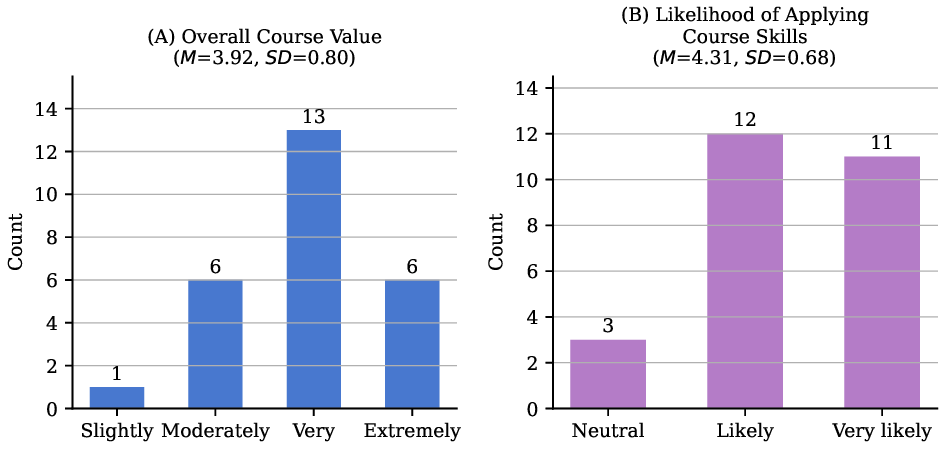}
  \caption{Overall course value and likelihood of applying course
  skills to future research ($n = 26$). Mean and SD are reported
  in each panel title.}
  \label{fig:value}
\end{figure}

Figure~\ref{fig:value} presents the distributions of overall course
value and likelihood of applying course skills to future research.
Seventy-three percent of students rated the course as very or extremely valuable ($M = 3.92$, $SD = 0.80$). Eighty-eight percent reported being likely or very likely to apply the skills learned to future research ($M = 4.31$, $SD = 0.68$), a transfer-intention indicator with strong implications for the course's design goal of developing durably applicable competencies. AI tool usefulness received high ratings ($M = 4.46$, $SD = 0.58$), with 96\% finding AI tools very or extremely useful for understanding complex research papers, suggesting that the course successfully framed AI tools as productive rather than threatening or irrelevant.

\FloatBarrier

\subsection{Qualitative Themes from Open-Response Items}

Open-response items asked students to identify the most valuable skill or technique they learned (Q21) and aspects of the course that could be improved (Q22). Responses were reviewed for recurring themes; direct quotations are paraphrased below to preserve student confidentiality given the small cohort.\\
Four themes dominated student responses. The most frequently cited
was \textit{verification and hallucination detection}: multiple
students identified learning to critically audit AI outputs and verify claims against original sources as the course's primary contribution, with one student describing this as developing ``the habit of critically evaluating AI-generated content by cross-checking sources.'' A second theme was \textit{prompt engineering and tool fluency}: students valued gaining systematic methods for designing effective prompts and learning to use AI platforms they had not previously encountered, including academic-specific tools. A third theme was \textit{literature organization and taxonomy construction}: several students cited the ability to use AI to organize large volumes of papers into structured frameworks as a skill directly applicable to their thesis and dissertation work. A fourth theme, mentioned by several respondents, was \textit{research gap identification}: students reported that the course gave them a structured method for identifying what is unstudied in their fields, a competency they described as previously tacit and uncodified. One student's response captured the integrative character of these gains: ``I learned how to use AI as a tool for structuring and refining my thinking rather than generating it.''\\
Student feedback on course improvement centered on two actionable
themes. The most consistent concern was \textit{assignment workload}: multiple students characterized weekly assignments as overly long and time-intensive, with some noting that comparable learning outcomes could have been achieved through more concise exercises. Several students also noted that in-class activities and homework felt redundant in some weeks. A second theme involved \textit{submission logistics and organization}: students requested more centralized submission through the course management system rather than across multiple platforms, and more consistent advance communication about weekly expectations. These feedback themes are directly relevant to the design limitation discussed in Section~\ref{sec:discussion}: future iterations should reduce weekly assignment length without sacrificing the documentation requirements that constitute the course's process-assessment architecture.

\subsection{Interpretation}

Taken together, the pre--post confidence data, learning outcome
ratings, and qualitative feedback present a consistent picture:
students perceived the course as developing the competencies it was
designed to develop, with the largest gains occurring precisely in
the domains most heavily scaffolded by the course's design. The very large effect sizes for AI attribution practice ($d = +2.40$) and hallucination detection ($d = +1.45$) are particularly noteworthy because these are the competencies most directly linked to the course's central theoretical commitments: the ethics-as-subject infrastructure and the experience-before-taxonomy sequencing of verification content, respectively.\\
The pattern of moderate, non-significant gains for the more general
research reading competencies is interpretively ambiguous but not
discouraging. A plausible reading is that these competencies, on
which students entered with relatively higher confidence, showed
genuine improvement that fell short of statistical significance at
the available sample size, rather than that the course failed to
develop them. This interpretation is supported by the uniformly
high post-course agreement on the learning outcome items
corresponding to these competencies.\\
These findings should be interpreted with appropriate caution.
Self-reported confidence is a proxy for, not a direct measure of,
actual competency development. The absence of a control condition
means that maturation, regression to the mean, or general exposure
to AI tools during the semester cannot be ruled out as partial
explanations. The sample size, while appropriate for a single course offering, limits statistical power and generalizability. These limitations motivate the ongoing formal analysis and planned
comparative study. What the present data establish is that the
course's design produced self-reported outcomes consistent with its
theoretical intentions, across a heterogeneous cohort, in its
inaugural offering.

\section{Discussion}
\label{sec:discussion}

\subsection{Contributions to AI Literacy Curriculum Design}

BSTA~495/395 makes several contributions to the emerging field of AI literacy curriculum design. First, it demonstrates that a
prerequisite-free, discipline-agnostic course can develop substantive AI research competencies at both undergraduate and graduate levels within a single semester, using the authentic task of literature review as the organizing spine of instruction rather than AI technology itself. Second, it provides a modular architecture (with a coherent design logic at each level, cumulative input design connecting modules, and explicit AI role framing at each stage) that other institutions can adapt to their own disciplinary contexts, course lengths, and student populations. Third, it offers an ethics infrastructure that treats AI integrity as a subject of instruction rather than a policy constraint, a distinction with meaningful implications for the durability and transferability of ethical practice.\\
The present course joins a small but growing set of documented,
prerequisite-free AI literacy curricula with sufficient detail to support replication. The closest comparable design is
\cite{shehu2026understanding}, whose UNIV~182 course demonstrates that non-major undergraduates can reach Bloom's \textit{Create} level through a technically-demanding pipeline of classifier construction and LLM evaluation. The two courses occupy different and complementary positions in the design space: UNIV~182 is organized around AI system construction and technical depth; BSTA~495/395 is organized around the cognitive demands of academic literature review and the scholarly competencies that AI-assisted research requires. A student who completes both would have covered the full arc from building AI systems to using them with the critical rigor that original research demands. The emergence of multiple such courses, each targeting a distinct workflow and learner population, reflects the maturation of AI literacy as a curricular genre and the recognition that no single course design can serve all of the contexts in which AI research literacy is now required.\\
The preliminary outcome evidence reported in
Section~\ref{sec:outcomes} is consistent with these design claims.
The very large effect sizes for the competencies most directly
targeted by specific design elements, particularly AI attribution
practice and hallucination detection, suggest that the design's
theoretical logic is at least partially validated by student
experience in the inaugural offering.

\subsection{Reconceptualizing Research Methods Education}

The course suggests a productive reconceptualization of research
methods education in the AI era. Traditional research methods courses emphasize methodological knowledge (understanding study designs, statistical approaches, and literature search strategies) without integrating the tool competencies students will use in practice. BSTA~495/395 inverts this relationship: it centers tool use while making underlying methodological knowledge visible as the evaluative standard against which AI outputs are assessed. Students acquire methodological knowledge in the process of verifying AI claims about methodology, a form of problem-based learning that may produce more durable understanding than didactic presentation alone.\\
Whether this instructional inversion produces superior methodological knowledge acquisition compared to traditional approaches is an empirical question that the ongoing outcome analysis will partially address. It is worth noting, however, that the question itself may be less important than it initially appears. Students in contemporary research environments will use AI tools regardless of whether their methods courses address those tools. A course that develops both methodological knowledge and the critical competencies to deploy AI tools in its service is arguably better preparation for the actual conditions of contemporary research practice than a course that develops one without the other.

\subsection{Limitations}

The course was developed and delivered by a single instructor in its inaugural offering at a single institution, limiting the evidence base for claims about effectiveness. Section~\ref{sec:outcomes} presents preliminary self-report outcome data from the Spring 2026 cohort; these findings are consistent with the design's intended effects but should be interpreted cautiously given the single-cohort, pre--post design without a control condition. Formal inferential analysis and a dedicated outcomes study with comparison data are ongoing.\\
The 13-week format, while achievable within a standard academic
calendar, is likely insufficient to fully develop the advanced
synthesis competencies addressed in Module D. Students with limited
prior research experience may require more practice time with full
literature review production than three weeks allows. Future
iterations may extend Module D or introduce the course as part of a
two-semester sequence for students pursuing thesis or independent
research projects.\\
The course also lacks explicit instruction in quantitative synthesis methods (meta-analysis, systematic review protocols, and PRISMA procedures) that complement the narrative and tabular synthesis methods addressed in Module D. Integration of these methods within the AI-assisted workflow is an important direction for course extension, particularly for students in health sciences, psychology, and other fields where systematic review is a standard product type.\\
Student feedback in the inaugural offering identified weekly
assignment length as the most consistent improvement concern. Future offerings should reduce per-assignment length while preserving the documentation requirements that constitute the course's process-assessment architecture, potentially by consolidating some weekly documentation into bi-weekly or module-level submissions.\\
Finally, the five platforms used for instruction represent a snapshot of the current tool landscape. The tool-agnostic design principle is intended to ensure that the competencies students develop remain applicable as specific platforms evolve or are superseded, but the course will require ongoing curation to ensure that illustrative platform examples remain current and representative.

\section{Conclusion and Recommendations}
\label{sec:conclusion}

BSTA~495/395 demonstrates that AI research literacy can be developed as a coherent, teachable competency within standard higher education structures, delivered to mixed-level, multi-disciplinary cohorts without prerequisites, and organized around the authentic cognitive demands of literature review rather than abstract AI principles. Its four-module architecture, scaffolded skill progression, ethics-as-subject infrastructure, and process-centered assessment system constitute a replicable design model for institutions seeking to develop comparable offerings. Preliminary evidence from the inaugural offering is consistent with the design's intended outcomes, with the largest self-reported gains occurring in the competency domains most directly targeted by specific instructional design elements. Five design principles, each grounded in the theoretical framework and supported by implementation observation, warrant particular emphasis for institutions developing AI research literacy
curricula.\\
\textbf{Design for transfer, not tools.} The AI landscape changes
too rapidly for platform-specific training to retain value. Curricula should develop transferable competencies (prompt construction, verification discipline, attribution practice, and responsible reliance judgment) that apply across tool generations and enable students to evaluate tools they have never encountered.\\
\textbf{Organize instruction around the research workflow.}
Effective AI research literacy requires understanding what AI does
and fails to do at each stage of literature review, not merely how
to operate individual platforms. Instruction organized around the
cognitive demands of the research process naturally produces the
stage-specific critical skills that responsible AI use requires.\\
\textbf{Sequence failure-mode instruction after tool experience.}
AI hallucinations and biases are better understood as organizational frameworks for personal experience than as abstract precautionary content. Positioning this instruction after students have accumulated meaningful AI tool experience produces deeper, more durable understanding of the verification practices it motivates.\\
\textbf{Treat ethics as subject, not policy.} Students who develop
principled understanding of why AI integrity requirements matter
carry that understanding across contexts; students who comply with
policies carry only the rules. The difference is not cosmetic:
principled understanding is what enables novel ethical judgment when students encounter AI use situations that no policy has anticipated.\\
\textbf{Assess process, not only product.} Weekly AI use
documentation, verification logs, and authorial mastery assessment
evaluate the process that produced student work, not only the quality of the work produced. In a course whose central goal is developing responsible research practice, process-level assessment is not supplementary; it is the primary instrument through which the course's core competencies are made visible and evaluable.\\
The students who complete this course leave with a replicable
workflow for AI-assisted research, the critical habits of mind to
apply it responsibly across a changing tool landscape, and the
documented experience to represent their AI use transparently in
professional and scholarly contexts. These are among the most
transferable and durable competencies that higher education can
develop in the current period of rapid technological transition.

\section*{Acknowledgments}
The author thanks the Office of the Provost, Lehigh University and
the H.S. Lee Family Foundation, Inc.\ for the AI Course
Transformation Grant, and Nesreen Haddush (Teaching Assistant,
BSTA~495/395, Spring 2026) for contributions to course implementation and student support, and the students in the Spring 2026 cohort whose engagement with the course materials informed the implementation observations reported in Section~\ref{sec:lessons} and the outcome findings reported in Section~\ref{sec:outcomes}.

\bibliographystyle{unsrt}
\bibliography{bmc_article}

\end{document}